\documentclass[pra,twocolumn,showpacs]{revtex4}
\usepackage{epsfig}
\begin{document}
%

%








%
\title{Probing Dynamics of Single Molecules: Non-linear Spectroscopy Approach}
\author{F. Shikerman, E. Barkai\\
Department of Physics, Bar Ilan University, Ramat-Gan 52900 Israel }
\begin{abstract}
{
 A two level model of a single molecule undergoing spectral diffusion dynamics
and interacting with a sequence of two short laser pulses is
investigated. Analytical solution for the probability of $n=0,1,2$
photon emission events for the telegraph and Gaussian processes are
obtained. We examine under what circumstances the photon statistics
emerging from such pump-probe set up provides new information on the
stochastic process parameters, and what are the measurement
limitations of this technique. The impulsive and selective limits,
the semiclassical approximation, and the fast modulation limit,
exhibit general behaviors of this new type of spectroscopy. We show,
that in the fast modulation limit, where one has to use impulsive
pulses in order to obtain meaningful results, the information on the
photon statistics is contained in the molecule's dipole correlation
function, equivalently to continuous wave experiments. In contrast,
the photon statistics obtained within the selective limit depends on
the both spectral shifts and rates and exhibits oscillations, which
are not found in the corresponding line-shape.}
\end{abstract}
\maketitle
%
%
\setcounter{equation}{0}
\section{Introduction}

Recently van Dijk et al \cite{vanDijk} reported the first
experimental ultra-fast pump-probe study of a single molecule
system. Unlike previous approaches to non-linear spectroscopy where
only the ensemble average response to the external fields is
resolved \cite{MukamelB}, the new method yields direct information
on single molecule dynamics, gained through the analysis of photon
statistics. Although the original experiment \cite{vanDijk} was
conducted on a molecule undergoing a relatively simple relaxation
process, the potential of combining non-linear spectroscopy with
single molecule spectroscopy inspires many unanswered questions:
What are the limitations of the investigation of fast dynamics? How
does the information contained in these experiments differ from the
information contained in simpler continuous wave experiments? How to
design the external control fields, so that needed information on
dynamics of molecules is gained? What is the finger print of
coherence in these types of experiments, and how its influence on
photon statistics is suppressed due to dephasing processes? The
answers to these questions are important for better understanding of
a wide variety of physical phenomena and have implication in the
investigation of ultra-fast dynamics of molecules in the condensed
phase, of quantum properties of light, and in the field of quantum
information and computation
\cite{vanDijk,Santori,Knill,Shih,Bouwmeester,Blatt,Katz,Orrit}. Here
we present a treatment based on the stochastic Kubo-Anderson model
\cite{MukamelB,Kubo,Tanimura,SB}, which yields general insights on
the
problem.\\

  We consider a sequence of two identical laser pulses interacting with a
single molecule (or an atom, or a quantum dot) undergoing a spectral
diffusion process, namely a molecule whose absorption frequency is
randomly modulated in time due to interaction with a thermal bath.
The electronic states of the single emitter are modeled based on the
two level system approximation. Most single molecules have a triplet
state, however the life time of the triplet is much longer than the
time scales under consideration in this paper, and it can be
neglected. It is assumed that the pulses are very short compared
with the inverse rate $R$ of the spectral diffusion process, as well
as with the inverse of the radiative life-time of the emitter,
$\Gamma$. The probability of photon emissions during the pulse event
is then negligible, and a pair of pulses yields two photons at most.
Repeating the experiment many times one obtains the probabilities
$\langle P_0\rangle, \langle P_1\rangle$ and $\langle P_2\rangle$ of
emitting $0, 1$ and $2$ photons, where $\langle\cdots\rangle$
designates the average over the stochastic modulation of the two
level system's absorption frequency $\omega(t)$. In what follows we
generalize the results obtained in our earlier publication \cite{SB}
by: (i) establishing the general expressions for $\langle
P_0\rangle, \langle P_1\rangle$ and $\langle P_2\rangle$ in the
limit of long measurement times without any restricting assumptions
regarding the laser detuning, (ii) comparing the photon statistics
obtained for the two state Kubo-Anderson and Gaussian processes. We
show, that under certain conditions this type of photon statistics
reveals important information on single molecule dynamics,
information which might be difficult to obtain using other
theoretical approaches to single molecule spectroscopy
\cite{BarkaiPRL,BarkaiRev,Brown,Mukamel,Xie,Goppich,Geva,YongPRL,Sanda}.\\

The general expressions for the photon statistics are obtained
starting with the path interpretation of Mollow and Zoller, Marte
and Walls \cite{Mollow,Zoller} of the optical Bloch equations
\cite{CT}. We show, that depending on the characteristics of the
stochastic dynamics and the laser field parameters, different types
of non-linear spectroscopies emerge. In particular, sensitivity to
the phase accumulated by the system in the delay interval between
the pulses is found, and impulsive and selective type of
spectroscopies are considered in detail. The Kubo-Anderson spectral
diffusion process \cite{Kubo} used in this work is found in many
molecular systems \cite{BarkaiRev,YongPRL,Geva,Sanda} and may be
easily detected, when the process is slow by means of the spectral
trail technique. Our goal is developing general methods suitable for
detection of the wider range of dynamics. Some technical details of
the calculations skipped in the main text appear in Appendixes A, B
and C.

%
%
\section{Photon Statistics}
\setcounter{equation}{0}
\begin{figure}
\includegraphics[width=\columnwidth]{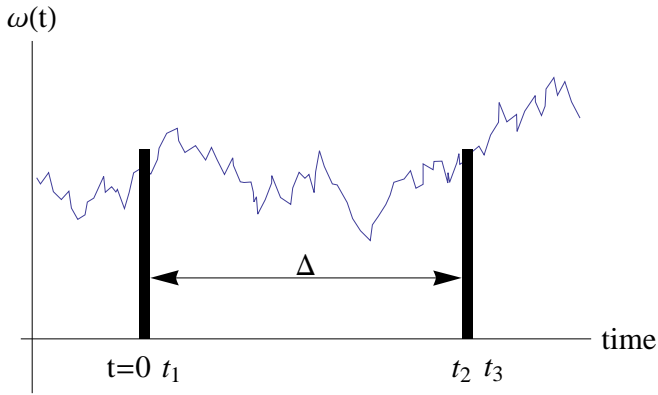}\\
  \caption{The two square laser pulses as modeled in Eq.~(\ref{eqft}) on the background of a schematic realization
of spectral diffusion process $\omega(t)$. $T=t_1=t_3-t_2$ is the
width of single pulse and $\Delta$ is the delay time between two
subsequent pulses. }\label{model1}
\end{figure}

  In our model two identical square pulses interact with the
two level system (see Fig.~1). The state of the system is described
by the density matrix, represented by the 4-vector
$\sigma=(\sigma_{\rm ee}, \sigma_{\rm gg}, \sigma_{\rm ge},
\sigma_{\rm eg})^{T}$. Here $\sigma_{\rm ee}$ and $\sigma_{\rm gg}$
represent the populations of the excited and ground states
respectively and $\sigma_{\rm ge},\sigma_{\rm eg}$ describe the
coherences, namely the off diagonal matrix elements of the density
matrix. For the sake of mathematical convenience we use the
following 4-dimensional basis \cite{SHB}: $| {\rm e}
\rangle=(1,0,0,0)^T$, which means that the system is in the pure
excited state, $|{\rm g} \rangle= (0,1,0,0)^T$ - the ground state,
and $| {\rm c} \rangle=(0,0,1,0)^T$ and $| {\rm c}^{*} \rangle=
(0,0,0,1)^T$ describe the coherences. The dynamical evolution of the
density matrix in the presence of the external laser field is given
by the optical Bloch equation \cite{CT}
\begin{equation} \dot{\sigma} = L\left( t \right) \sigma +
\hat{\Gamma}\sigma. \label{eqBloch}
\end{equation}
The operator
\begin{equation}
L(t) = \left(
\begin{array}{c c c c}
- \Gamma  & 0 & - i \Omega f(t)  & i \Omega f(t) \\
0 & 0 & i \Omega f(t) & - i \Omega f(t) \\
- i \Omega f(t) & i \Omega f(t) & i \omega(t) - \Gamma/2 & 0 \\
 i \Omega f(t) & - i \Omega f(t) & 0 & - i \omega(t) - \Gamma/2
\end{array}
\right) \label{eqOBE}
\end{equation}
describes the interaction of the system with the driving
electromagnetic field through $\Omega f(t)$, where $
\Omega=-(1/\hbar){\bf d} _{\rm ge} \cdot {\bf E} _0 $ is the Rabi
frequency with ${\bf E_0}$ - the amplitude of the electric field and
${\bf d} _{\rm ge}={\bf d} _{\rm eg}$ the transition dipole moment
of the two level system. The operator
$\hat{\Gamma} =\Gamma |{\rm g}  \rangle \langle {\rm e}| $
describes the transition from the excited state into the ground
state, due to spontaneous  emission with $\Gamma$ designating the
emission rate. Finally $\omega (t)$ is the stochastic time dependent
absorption frequency of the system. The spectral diffusion process
$\omega(t)$ is modeled using the Kubo-Anderson approach:
\begin{equation}\omega(t)=\omega_0 + \delta w(t),\end{equation}
where $\omega_0$ is the bare absorption frequency of the single
emitter, and $\delta w(t)$ is a random function of time
\cite{Kubo,MukamelB}. We will assume the process $\delta w(t)$ is
stationary, its mean is zero, its correlation function is
\begin{equation}\langle \delta w(t_0+t) \delta w(t_0) \rangle=\nu^2
\psi(t),\label{corf}\end{equation}
where $\psi(0)=1$  and $\psi(\infty)=0$. Later we will demonstrate
our results for a particular choice
\begin{equation}\psi(t)=e^{-2Rt}\label{psi},\end{equation}
obtaining semi-analytical solution for the Gaussian process and
analytical solution for the two-state Kubo-Anderson process
\cite{BarkaiRev,Berez}, where $\omega(t)=\omega_0 + \nu$ or
$\omega(t)=\omega_0 - \nu$, with the rate $R$ determining the
transition between the $+$ and $-$ states. The later is used to
model single molecules in low temperature glasses \cite{Geva}.\\

In the case of two identical square pulses the modulating function
$f(t)$ in Eq.~(\ref{eqOBE}) is:
\begin{equation}
f(t)= \left\{
\begin{array}{l l}
\cos \left( \omega_L t \right) & 0<t<t_1 \\
0 & t_1<t<t_2 \\
\cos \left[ \omega_L( t -t_2) \right] & t_2<t<t_3 \\
0  & t_3<t
\end{array}
,\right. \label{eqft}
\end{equation}
where $\omega_L$ is the laser frequency, $t_1=t_3-t_2=T$ is the
pulses duration, and $\Delta=t_2 - t_1$ is the delay between the
pulses. In our calculations we assume: (i) that the system is always
found in its ground state at the beginning of the experiment. (ii)
In order to get meaningful measurable results one has to use
sufficiently intense laser fields such that $\Omega T\sim 1$ - weak
fields cannot excite the emitter even once. (iii) The pulses are
short enough, so that the lifetime of the excited state is much
longer than pulse's duration - $\Gamma T\ll 1$ (hence,
$\Omega\gg\Gamma$). (iv) The rate $R$ of changes of the stochastic
process $\omega(t)$ satisfies $R T \ll 1$ (hence, $\Omega\gg R$).
Assumption (iii) leads to the mentioned negligibility of photon
emissions during the pulse events. The last assumption implies, that
the time-dependent absorption frequency $\omega(t)$ is unchanged
during the excitations, and will be taken to be
$\omega(t_1)=\omega_0+\delta w(t_1)$ during the first pulse event
and
$\omega(t_2)=\omega_0+\delta w(t_2)$ during the second pulse event.\\

A calculation given in short in Appendix A yields the following
expression for the probabilities of emitting zero, one and two
photons in the limit of long measurement times $t\to\infty$ for a
particular realization of the stochastic process $\omega(t)$:
\begin{widetext}
\begin{equation} P_n[\omega\left(t_1\right),\omega\left( t_2 \right),\theta (\Delta)]=P_n ^{{\rm Cla}} \left[\omega\left(t_1\right),\omega\left( t_2 \right)\right] +
2 e^{-\frac{\Gamma \Delta}{2}}{\rm Re}\left\{A_n^{{\rm Coh}}
\left[\omega\left(t_1\right),\omega\left( t_2 \right) \right]{\rm
e}^{ i (\theta(\Delta) +\omega_0(T+\Delta)) } \right\}, \label{PnSD}
\end{equation}
where $\theta(\Delta)$ is the random phase accumulated during the
delay interval given by
\begin{equation}
\theta(\Delta)=\int_{0} ^{\Delta} \delta w(t) {\rm d}
t,\label{theta}
\end{equation}
and $P_n ^{{\rm Cla}} \left[\omega\left(t_1\right),\omega\left( t_2
\right)\right]$ and $A_n ^{{\rm Coh}}
\left[\omega\left(t_1\right),\omega\left( t_2 \right)\right]$ are
given in Table~1, where ${\cal
G}\left[\omega\left(t_i\right)\right]$ $(i=1,2)$ is the Green
function defined in Eqs~(\ref{calG},\ref{eqRWA}) below. Note, from
Eq.~(\ref{PnSD}) it follows that all accumulated random phase
effects become negligible when $\Gamma\Delta\gg1$. The two summands
in Eq.~(\ref{PnSD}) describe two kinds of possible quantum
trajectories leading to required number of photon emission events:
the term $P_n ^{{\rm Cla}}\left[\omega\left(t_1\right),\omega\left(
t_2 \right)\right]$ we call semiclassical in the sense that it
summarizes the paths where at the beginning of the delay the system
is found in one of the pure states $|{\rm e}\rangle$ or $|{\rm
g}\rangle$ \cite{Remark}. For example, consider the first term of
$P_0^{{\rm Cla}}$: the system starts in the electronic ground state
$| {\rm g} \rangle$, then it evolves with the propagator of the
first pulse ${\cal G}\left[\omega\left(t_1 \right) \right]$ without
photon emissions and reaches the excited state $|{\rm e} \rangle$,
it stays in the excited state during the delay interval (with
probability $e^{-\Gamma \Delta}$), and afterwards the second pulse
with the propagator ${\cal G}\left[\omega\left(t_2 \right) \right]$
stimulates the induced emission, bringing the system back to the
ground state $| {\rm g}\rangle$ without emitting a photon. The
second summand in the right-hand side of Eq.~(\ref{PnSD}), $A_n
^{{\rm Coh}}\left[\omega\left(t_1\right),\omega\left( t_2
\right)\right]$, summarizes the contribution of the coherence
effects (i.e. all those quantum paths where at the beginning of the
delay interval the system is left in a superposition of the pure
states). It can be shown (see Appendix A), that very strong or
resonant $\pi$-pulses, where $\Omega T=\pi$, simply switch the state
of the molecule being in $|{\rm g}\rangle$ to $|{\rm e}\rangle$ or
vice versa. Hence, they do not excite the coherence - in such cases
the coherent term $A_n ^{{\rm
Coh}}\left[\omega\left(t_1\right),\omega\left( t_2 \right)\right]$
vanishes.
$$ \mbox{
\begin{tabular}{|c| c| c|}
\hline $n$ & $P_n ^{{\rm Cla}}
\left[\omega\left(t_1\right),\omega\left( t_2 \right)\right]$ &
$A_n^{{\rm Coh}} \left[ \omega\left(t_1\right),\omega\left( t_2 \right) \right]$ \\
\hline $0$ & $ \langle {\rm g}| {\cal G}\left[\omega\left(t_2
\right) \right] | {\rm e} \rangle \langle {\rm e} | {\cal
G}\left[\omega\left(t_1 \right) \right] | {\rm g}\rangle e^{ -
\Gamma \Delta} + \langle {\rm g}| {\cal G}\left[\omega\left(t_2
\right) \right]  | {\rm g} \rangle \langle {\rm g}|{\cal
G}\left[\omega\left(t_1 \right) \right] | {\rm g} \rangle  $ & $
\langle {\rm g} |
{\cal G}\left[\omega\left(t_2 \right) \right]  | {\rm c} \rangle \langle c |  {\cal G}\left[\omega\left(t_1 \right) \right] | {\rm g} \rangle $ \\
\ & \ & \\
\hline $ 1$ &  $ \langle {\rm g} | {\cal G}\left[\omega\left(t_2
\right) \right] | {\rm g} \rangle\langle {\rm e} | {\cal
G}\left[\omega\left(t_1 \right) \right] | {\rm g} \rangle + \langle
{\rm e} | {\cal G}\left[\omega\left(t_2 \right) \right] | {\rm g}
\rangle \langle {\rm g} |{\cal G}\left[\omega\left(t_1 \right)
\right] | {\rm g} \rangle$&
$ \langle {\rm e} | {\cal G}\left[\omega\left(t_2 \right) \right] |c\rangle \langle c |  {\cal G}\left[\omega\left(t_1 \right) \right] | {\rm g} \rangle $ \\
\ & \ & \\
\hline $ 2$ & $ \langle {\rm e} | {\cal G}\left[\omega\left(t_2
\right) \right]  | {\rm g} \rangle \langle {\rm e} | {\cal
G}\left[\omega\left(t_1 \right) \right] | {\rm g} \rangle\left(1 -
e^{ - \Gamma \Delta} \right) $ &
$ 0 $ \\
& \ & \  \\
\hline
\end{tabular}}$$
$$\mbox{\textbf{Table~1:} Photon statistics for two short pulses and arbitrary spectral diffusion process $\omega(t)$.}$$
\end{widetext}

In Table~1 the propagator of the two level system during the first
and the second pulse events ${\cal
G}\left[\omega\left(t_i\right)\right]$ $(i=1,2)$ is obtained within
the rotating wave approximation (RWA) (see Appendix A):
\begin{equation}{\cal G}\left[\omega\left(t_i\right)\right]=e^{T\cdot L^{\rm
RWA}[\omega(t_i)]},\label{calG}\end{equation}
where
\begin{equation}
L^{{\rm RWA}}[\omega(t_i)] = \left(
\begin{array}{c c c c}
0 & 0 & {-i \Omega \over 2} & {i \Omega \over 2} \\
0 & 0 & {i \Omega \over 2} & {-i \Omega \over 2} \\
 {-i \Omega \over 2} & {i \Omega \over 2} & - i \delta(t_i) & 0 \\
 {i \Omega \over 2} & {-i \Omega \over 2} & 0 &  i \delta(t_i)
\end{array}
\right), \label{eqRWA}
\end{equation}
and
\begin{equation}
\delta(t_i)=\delta_L- \delta w(t_i)\label{detuning}
\end{equation}
with \begin{equation}\delta_L
=\omega_L-\omega_0\label{detuningL}\end{equation}
is the detuning. The mathematical calculations were made with the
help of Mathematica 5.0. Evidently, since the spontaneous emissions
during the pulse events are neglected ($\Omega\gg\Gamma$), the
Green function Eq.~(\ref{calG}) describes well-known Rabi oscillations.\\

%
%
\section{Influence of Spectral Diffusion on Photon Statistics.}

We now take the average of $P_n[\omega\left(t_1\right),\omega\left(
t_2 \right),\theta (\Delta)]$ Eq.~(\ref{PnSD}) over the stochastic
process $\omega(t)$. This procedure requires the knowledge of the
joint probability density function (PDF) ${\cal
P}[\omega(t_1),\omega(t_2),\theta (\Delta)]$ of finding the system's
absorption frequency $\omega(t)$ in the infinitesimal range near
$\omega_0+\delta w(t_1)$ at $t_1$, near $\omega_0+\delta w(t_2)$ at
$t_2$,  with accumulated random phase $\theta( \Delta)$. Then by
definition the average of $P_n$ is:
$$
\langle P_n\rangle=\int_{-\infty}^{\infty}{\rm d
}\theta(\Delta)\int_{0}^{\infty}{\rm d
}\omega(t_1)\int_{0}^{\infty}{\rm d }\omega(t_2)\times$$
\begin{equation}
\times P_n[\omega(t_1),\omega(t_2),\theta (\Delta)]{\cal P}[\omega(t_1),\omega(t_2),\theta (\Delta)].\label{PnSDgen}\\
\end{equation}
Later we find exact solution for the three-variable PDF ${\cal
P}[\omega(t_1),\omega(t_2),\theta (\Delta)]$ for the case of
two-state Kubo-Anderson process, thus providing all essential tools
for calculation of $\langle P_n\rangle$ in the case of the telegraph
noise. But first, we discuss several
limiting cases common for all stationary processes.\\

\begin{figure}
 \includegraphics[height=1.8in, width=3.1in]{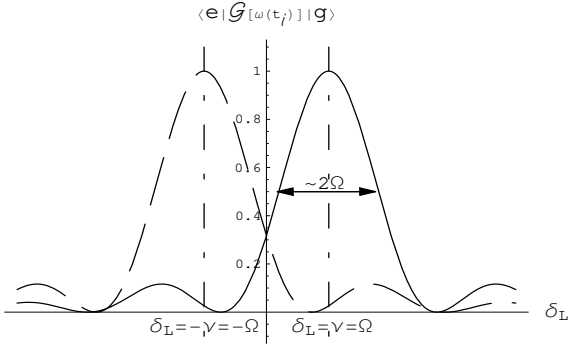}
\caption{ The matrix element $\langle {\rm e}|{\cal
G}[\omega(t_i)]|{\rm g}\rangle$ Eq.~(\ref{matrxelm}) describing the
probability of transition from the ground to excited state as a
result of interaction with a $\pi$-pulse for $\Omega=4\Gamma$. The
{\em smooth curve} represents $\langle {\rm e}|{\cal
G}[\omega(t_i)]|{\rm g}\rangle$ for $\omega(t)=\omega_0+\nu$ with
$\nu=\Omega$. The {\em dashed curve} represents $\langle {\rm
e}|{\cal G}[\omega(t_i)]|{\rm g}\rangle$ for
$\omega(t)=\omega_0-\nu$ with $\nu=\Omega$. The figure illustrates
that since the two curves practically do not overlap for
$\nu\geq\Omega$ the probabilities to excite the molecule being in
state $\omega(t_1)=\omega_0+\nu$ during the first pulse event and in
the state $\omega(t_2)=\omega_0-\nu$ during the second pulse event
are measurably different.}
\end{figure}

In Fig.~2 we plotted the probability of transition from the ground
to the excited state $\langle {\rm e}|{\cal G}[\omega(t_i)]|{\rm
g}\rangle$ Eq.~(\ref{matrxelm}) for two identical $\pi$-pulses. The
smooth and the dashed curves represent $\langle {\rm e}|{\cal
G}[\omega(t_i)]|{\rm g}\rangle$ as a function of $\delta_L$ for
$\omega(t)=\omega_0+\nu$ and $\omega(t)=\omega_0-\nu$ respectively
with $\nu=\Omega$. The half width of $\langle {\rm e}|{\cal
G}[\omega(t_i)]|{\rm g}\rangle$ is of the order of $\Omega$. When
$\nu\gg\Omega$ the two curves practically do not overlap. Hence,
assuming the absorption frequency of the system at the moment of the
first excitation is $\omega(t_1)=\omega_0+\nu$ and of the second
excitation is $\omega(t_2)=\omega_0-\nu$, the probabilities to
excite the molecule during two pulse events strictly differ. In this
selective limit the photon statistics is therefore very sensitive to
the temporal state of the molecule at the moments of excitations,
and the particular type of the underlying stochastic process has
large importance. Later we consider this limit in detail
for the case of telegraph noise. The opposite situation is the impulsive limit.\\

{\em Impulsive limit $\Omega\gg\nu$.~~-}~~In the limit
$\Omega\gg\nu$, which we call impulsive, the matrix elements of
${\cal G}[\omega(t_1)]$ and ${\cal G} [ \omega(t_2)]$
Eqs.~(\ref{matrxelm}) {\em become independent of the value of
stochastic detuning $\delta w(t)$ at the moment of the excitation}.
Thus instead of the multi variable PDF
$P[\omega(t_1),\omega(t_2),\theta (\Delta)]$ we now have to deal
only with the one variable PDF of the phase $\theta (\Delta)$. As a
result the photon statistics shows an interesting relation with
linear continuous wave spectroscopy:
\begin{equation}\lim_{\Omega\gg\nu}\langle P_n\rangle=\lim_{\Omega\gg\nu}P_n ^{\rm
Cla} + 2 e^{-\frac{\Gamma \Delta}{2}} {\rm Re}\left[ \phi(\Delta)
e^{ i\omega_0 (T+\Delta)} \lim_{\Omega\gg\nu}A_n^{\rm Coh}\right],
\label{PnImp}
\end{equation}
where using Table~1 and Eqs.~(\ref{matrxelm})
\begin{widetext}
\begin{equation}
\lim_{\Omega\gg\nu}\langle P_1^{\rm Cla}\rangle=\frac{2\Omega^2
\left[\delta_L^2+\Omega^2\cos^2\left(\frac{\Omega
T}{2}\sqrt{1+\frac{\delta_L^2}{\Omega^2}}\right)\right]\sin^2\left(\frac{\Omega
T}{2}\sqrt{1+\frac{\delta_L^2}{\Omega^2}}\right)}
{(\delta_L^2+\Omega^2)^2}\label{PnImp1}\end{equation}
\begin{equation}
\lim_{\Omega\gg\nu}\langle P_2^{\rm
Cla}\rangle=\frac{(1-e^{-\Gamma\Delta})}{(\delta_L^2+\Omega^2)^2}\Omega^4\sin^4\left(\frac{\Omega
T}{2}\sqrt{1+\frac{\delta_L^2}{\Omega^2}}\right),\label{PnImp2}\end{equation}
\begin{equation}
\lim_{\Omega\gg\nu}\langle P_0^{\rm
Cla}\rangle=1-\lim_{\Omega\gg\nu}\langle P_1^{\rm
Cla}\rangle-\lim_{\Omega\gg\nu}\langle P_2^{\rm
Cla}\rangle\label{PnImp0}\end{equation}
%
and
\begin{equation}
\lim_{\Omega\gg\nu}A^{\rm Coh}_0=\frac{\Omega^2\left(2\delta_L
\sin^2\left(\frac{\Omega
T}{2}\sqrt{1+\frac{\delta_L^2}{\Omega^2}}\right)+i\sqrt{\delta_L^2+\Omega^2}\sin
\left(\Omega T\sqrt{1+\frac{\delta_L^2}{\Omega^2}}\right)\right)^2}
{4\left(\delta_L^2+\Omega^2\right)^2}.\label{AnImp}\end{equation}
\end{widetext}
In Eq.~(\ref{PnImp}) the function $\phi(\Delta)$ given by
\begin{equation}
\phi(\Delta) =\langle\exp[i\theta(\Delta)]\rangle =\langle \exp[ i
\int_0 ^\Delta \delta \omega(t) {\rm d} t ]\rangle \label{phi}
\end{equation}
is the well investigated Kubo-Anderson correlation function, whose
Fourier transform is the line shape of the two level system
according to the Wiener-Khintchine theorem \cite{Kubo}.  In
conclusion, we see, that working with very strong laser fields
$\Omega\gg\nu$ under
assumptions (iii,iv) we gain the same information as found in the line-shape in continuous wave experiments.\\

  Near the resonance, where $\delta_L\sim 0$, using Eqs.~(\ref{PnImp}-\ref{AnImp}) we find
%
%
$$ \lim_{\Omega\gg\nu}  \langle P_0 \rangle  =
 e^{-\Gamma \Delta} \sin^4\left( { \Omega T \over 2} \right)  + \cos^4 \left( { \Omega
T \over 2} \right) -$$\begin{equation}-{1 \over 2} e^{ - \Gamma
\Delta/2} \sin^2 \left( \Omega T \right) {\rm Re}\left[ \phi(\Delta)
e^{ i \omega_0 (T+\Delta)} \right],\label{Eqpooo0}\end{equation}
\begin{equation}
\lim_{\Omega\gg\nu} \langle  P_1 \rangle = {1 \over 2} \sin^2
\left(\Omega T\right) \left\{ 1  +  e^{ - {\Gamma \Delta \over 2} }
{\rm Re} \left[\phi(\Delta) e^{ i \omega_0 (T+\Delta)}
\right]\right\}, \label{Eqpooo1}\end{equation}
\begin{equation}
\lim_{\Omega\gg\nu} \langle P_2 \rangle= \left(1 - e^{ - \Gamma
\Delta}\right) \sin^4\left( { \Omega T \over 2}\right).
\label{Eqpooo2}
\end{equation}
%
From Eqs.~(\ref{Eqpooo0}-\ref{Eqpooo1}) we see, that for the strong
$\pi$-pulses the coherent terms vanish, as mentioned. In contrast,
for $\pi/2$-pulses with $\Omega T=\pi/2$ the importance of the
coherent terms, and hence, the correlation function $\phi(\Delta)$
on the photon statistics is the strongest, since the $\pi/2$ pulse
excites the off diagonal terms of the pulse-propagators ${\cal
G}[\omega(t_i)]$ Eq.~(\ref{matrxelm}) \cite{SHB}.\\

{\em Semiclassical approximation.~~-}~~The influence of the
coherence on photon statistics in many experimental cases is
expected to be difficult to detect. It may be because of the
dephasing effects caused by the damping coefficient
$e^{-\frac{\Gamma \Delta}{2}}$ multiplying the coherent terms
$A_n^{\rm Coh}$ in Eq.~(\ref{PnSD}). Moreover, because of the large
value of the bare optical transition frequency $\omega_0$ the
coherent paths oscillate too fast to be detected (see the term
$e^{i\omega_0(T+\Delta)}$ in Eq.~(\ref{PnSD})). In such cases a
practical approximation is to keep only the semiclassical terms $P_n
^{{\rm Cla}}$. Nevertheless, we stress, that for multilevel systems
or in non optical experiments on Josephson junction coherence
contribution is important \cite{Katz}. Since the semiclassical paths
are independent of the random phase $\theta(\Delta)$, for the
calculation of $\langle P_n^{\rm Cla}\rangle$ we need only the
marginal, two dimensional PDF ${\cal P}[\omega(t_1),\omega(t_2)]$.
For example, in the case of Gaussian noise \cite{Risken} we have:
$$
{\cal P}[\delta w(t_1),\delta
w(t_2)]=\frac{1}{2\pi\nu^2\sqrt{(1-\psi^2)}}\times$$
\begin{equation}
\times \exp\left[-\frac{\delta w(t_1)^2+\delta w(t_2)^2-2\delta
w(t_1)\delta w(t_2)\psi}{2\nu^2(1-\psi^2)}\right], \label{2P}
\end{equation}
where $\psi=\psi(|t_2-t_1|)$ is the time dependent part of the
correlation function in Eq.~(\ref{corf}). Once the two-time PDF
${\cal P}[\omega(t_1),\omega(t_2)]$ is known, $\langle P_n\rangle$
within the semiclassical approximation is:
\begin{equation}
\!\!\!\!\langle P_n ^{{\rm Cla}} \rangle\!\!= \!\!\int_0 ^\infty
\!\!\!\!\int_0 ^\infty \!\!\!\!\!P_n ^{{\rm
Cla}}[\omega(t_1),\omega(t_2)] {\cal P}[\omega(t_1),\omega(t_2)]
{\rm d} \omega(t_1) {\rm d} \omega(t_2).
\end{equation}
In Appendix B we give the explicit semiclassical approximation for
the two state Kubo-Anderson model Eqs.~(\ref{p0av}-\ref{p2av}).
Later in Fig.~3 we compare these results with similar results for
the Gaussian noise, which was solved semi-analytically with the help
of Mathematica 5.0. In the both calculations the same correlation function Eqs.~(\ref{corf},\ref{psi}) was used.\\

\section{Two State Process: Exact solution}

Now we obtain the exact solution for the two state Kubo-Anderson
Poissonian process, where the absorption frequency of the system
jumps between the $+$ and $-$ states, i.e. $\omega(t)=\omega_0\pm
\nu$. We denote the initial state, during the first pulse with
$\omega(t_1)=+$ or $\omega(t_1)=-$. Similarly, the final state at
the second pulse is $\omega(t_2)=+$ or $-$. Since the random phase
is now given by $\Delta \theta = \nu( T^{+} - T^{-})$, where
$T^{\pm}$ are occupation times in states $+$ and $-$ \cite{Berez}
obeying $\Delta =T^{+} + T^{-}$, the joint PDF ${\cal
P}[\omega(t_1),\omega(t_2),\theta(\Delta)]$ can be found from the
joint PDF $h[\omega(t_1),\omega(t_2),T^{+}]$ of finding the system
in state $\omega(t_1)=\pm$ during the first pulse, state
$\omega(t_2)=\pm$ during the second and with the occupation time
$T^{+}$ between the two pulses. In this case, where the random
process takes only discrete values, $\langle P_n\rangle$
Eq.~(\ref{PnSDgen}) takes the form
\begin{widetext}
\begin{equation}\!\!\!\!\!\!\!\!\!\langle P_n \rangle = \!\!\!\!\!\!\!\!\!\! \!\!\sum_{\omega(t_1),\omega(t_2)=\pm}\!\!\!\!\!\!\!\! \!\!{\cal P}[\omega(t_1),\omega(t_2)] P_n ^{{\rm
Cla}} \left[\omega(t_1),\omega(t_2)\right]\:+\:2{\rm Re}\left\{ {\rm
e}^{ i \omega_0(T+\Delta)- \frac{\Gamma \Delta}{2} }
 A_n^{{\rm Coh}} \left[
\omega(t_1),\omega(t_2)\right] \int_{0}^{\infty}e^{i\nu( 2T^{+} -
\Delta)}h\left[\omega(t_1),\omega(t_2),T^{+}\right]\,dT^{+}\right\},
\label{PnSDAv}
\end{equation}
where
\begin{equation}
{\cal P}[\pm ,\pm]=\frac{1+e^{- 2 R \Delta}}{4}, ~~~ {\cal P}[\pm,
\mp]=\frac{1-e^{- 2 R \Delta}}{4}\label{Pif}\end{equation}
are the probabilities of finding the particle initially in state
$\omega(t_1)=\pm$ and finally in state $\omega(t_2)=\pm$, which are
easy to obtain from Poissonian statistics. In the integrand of
Eq.~(\ref{PnSDAv}) one can recognize Laplace transform $T^{+} \to -2
i \nu$ of $h\left[\omega(t_1),\omega(t_2),T^{+}\right]$. Simple
rearrangement leads to:
\begin{equation}
\langle P_n \rangle = \!\!\!\!
\!\!\sum_{\omega(t_1),\omega(t_2)=\pm}\!\!\!\! \!\! {\cal
P}[\omega(t_1),\omega(t_2)] P_n ^{{\rm
Cla}}[\omega(t_1),\omega(t_2)] +2{\rm Re}\left\{e^{- \frac{\Gamma
\Delta}{2}} e^{i \left( \omega_0 - \nu\right)\Delta+i\omega_0
T}\hat{h}\left[\omega(t_1),\omega(t_2),- 2 i \nu \right] A_n^{{\rm
Coh}} [\omega(t_1),\omega(t_2)] \right\}, \label{eqPn11}
\end{equation}
where $\hat{h}\left[\omega(t_1),\omega(t_2),- 2 i \nu \right]$ is
the Laplace $T^{+} \to -2 i \nu$ transform of
$h[\omega(t_1),\omega(t_2),T^{+}]$. The procedure of derivation of
$\hat{h}\left[\omega(t_1),\omega(t_2),- 2 i \nu \right]$, based on
the renewal processes theory \cite{Berez, Luck}, is given in detail
in Appendix C. Here we present the final results:
\begin{equation}
\hat{h}\left[\pm,\pm, - 2 i \nu\right] = {e^{- \Delta(R - i \nu)}
\over 2} \left[ \cosh\left( \Delta \sqrt{ R^2 - \nu^2} \right) \pm {
i \nu \sinh \left( \Delta \sqrt{ R^2 - \nu^2} \right) \over \sqrt{
R^2 - \nu^2} } \right], \label{eqPn13F1}\end{equation}
\begin{equation}
\hat{h}\left[\mp,\pm,-2 i \nu\right] = e^{ - \Delta( R - i \nu) } {
\sinh\left[ \Delta \sqrt{ R^2 - \nu^2} \right] R \over 2 \sqrt{ R^2
- \nu^2}}.
 \label{eqPn13F2}
\end{equation}
%
%
Using Eqs.~(\ref{Pif}-\ref{eqPn13F2},\ref{matrxelm}) and
Table~1 the calculation of $\langle P_n\rangle$ is straightforward.\\


{\em Semiclassical selective limit.--}  
Now we consider selective limit for the two state Kubo-Anderson
process within the semiclassical approximation. From the exact
solution Eqs.~(\ref{p1av}-\ref{p0av}), considering two opposite
situations $\Omega\sim\delta_L+\nu$ and $ \Omega\ll\delta_L-\nu$ or
$\Omega\sim\delta_L-\nu$ and $ \Omega\ll\delta_L+\nu$ we find:
%
$$
\langle \lim_{\nu\gg\Omega}P_{1}^{{\rm
Cla}}\rangle=\frac{(1+e^{-2R\Delta})\Omega^2\sin^2\left[\frac{\Omega
T}{2}\sqrt{1+\frac{(|\delta_L|-\nu)^2}{\Omega^2}}\right]}{2\left[(|\delta_L|-\nu)^2+\Omega^2\right]}\times\frac{(|\delta_L|-\nu)^2+\Omega^2\cos^2\left[\frac{\Omega
T}{2}\sqrt{1+\frac{(|\delta_L|-\nu)^2}{\Omega^2}}\right]}{(|\delta_L|-\nu)^2+\Omega^2}+$$
\begin{equation}
+\frac{(1-e^{-2R\Delta})\Omega^2\sin^2\left[\frac{\Omega
T}{2}\sqrt{1+\frac{(|\delta_L|-\nu)^2}{\Omega^2}}\right]}{2\left[(|\delta_L|-\nu)^2+\Omega^2\right]},
\label{p1avsel}
\end{equation}
\begin{equation}
\langle \lim_{\nu\gg\Omega}P_{2}^{{\rm
Cla}}\rangle=\frac{(1+e^{-2R\Delta})(1-e^{-\Gamma\Delta})\Omega^4\sin^4\left[\frac{\Omega
T}{2}\sqrt{1+\frac{(|\delta_L|-\nu)^2}{\Omega^2}}\right]}{4\left([(|\delta_L|-\nu)^2+\Omega^2\right]^2},
\label{p2avsel}
\end{equation}
%
%
\begin{equation}
\langle \lim_{\nu\gg\Omega}P_{0}^{{\rm Cla}}\rangle=1-\langle
\lim_{\nu\gg\Omega}P_{1}^{{\rm Cla}}\rangle-\langle
\lim_{\nu\gg\Omega}P_{2}^{{\rm Cla}}\rangle. \label{p0avsel}
\end{equation}
Notice, that $P_1^{\rm Cla}$ depends on $\Delta$ only trough $e^{-2R\Delta}$. Hence, using Eq.~(\ref{p1avsel}) it is easy to measure the dynamics of the molecule.\\

%

\begin{figure}
\begin{eqnarray*}
 \includegraphics[height=1.4in, width=2in]{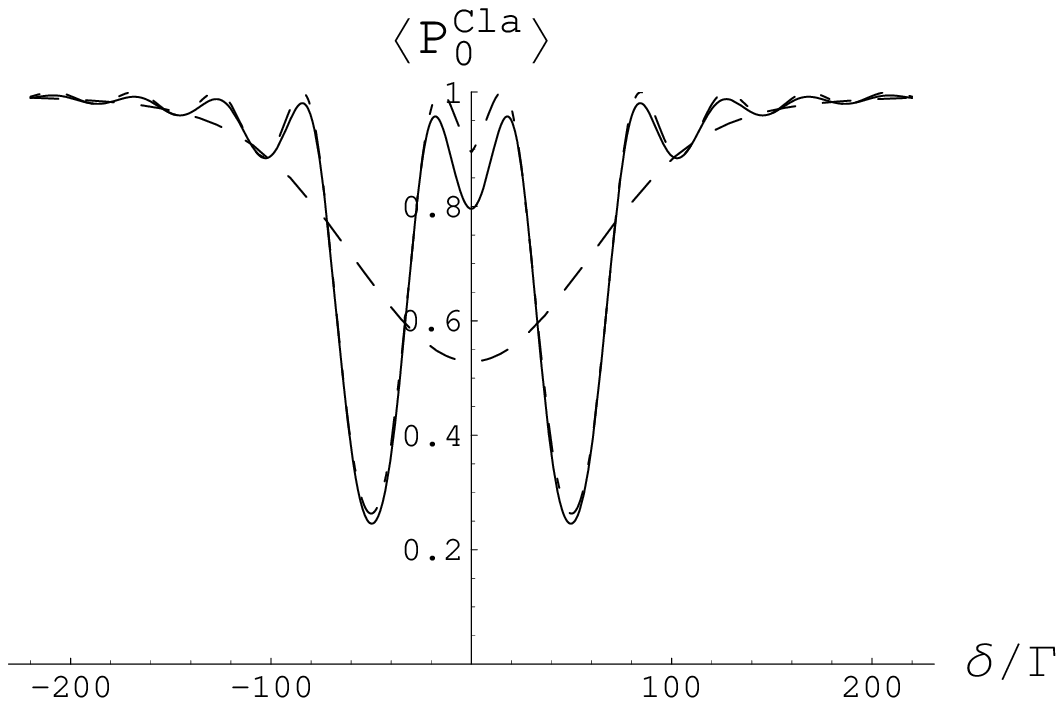}\textbf{a.}&
 \includegraphics[height=1.4in, width=2in]{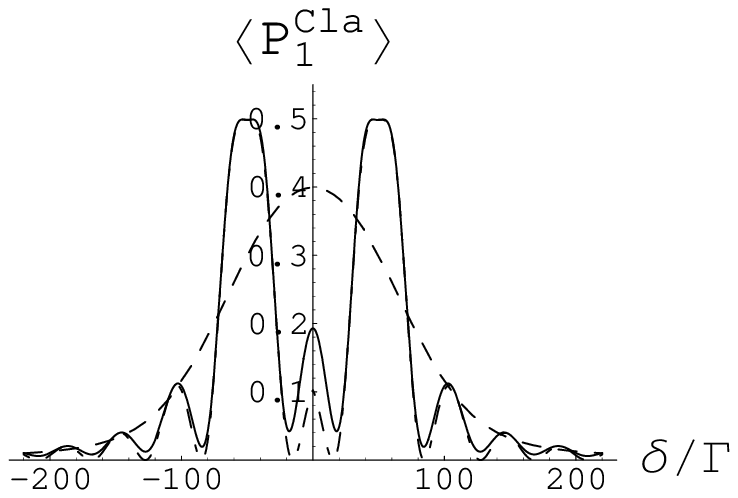}\textbf{b.}&
\includegraphics[height=1.4in, width=2in]{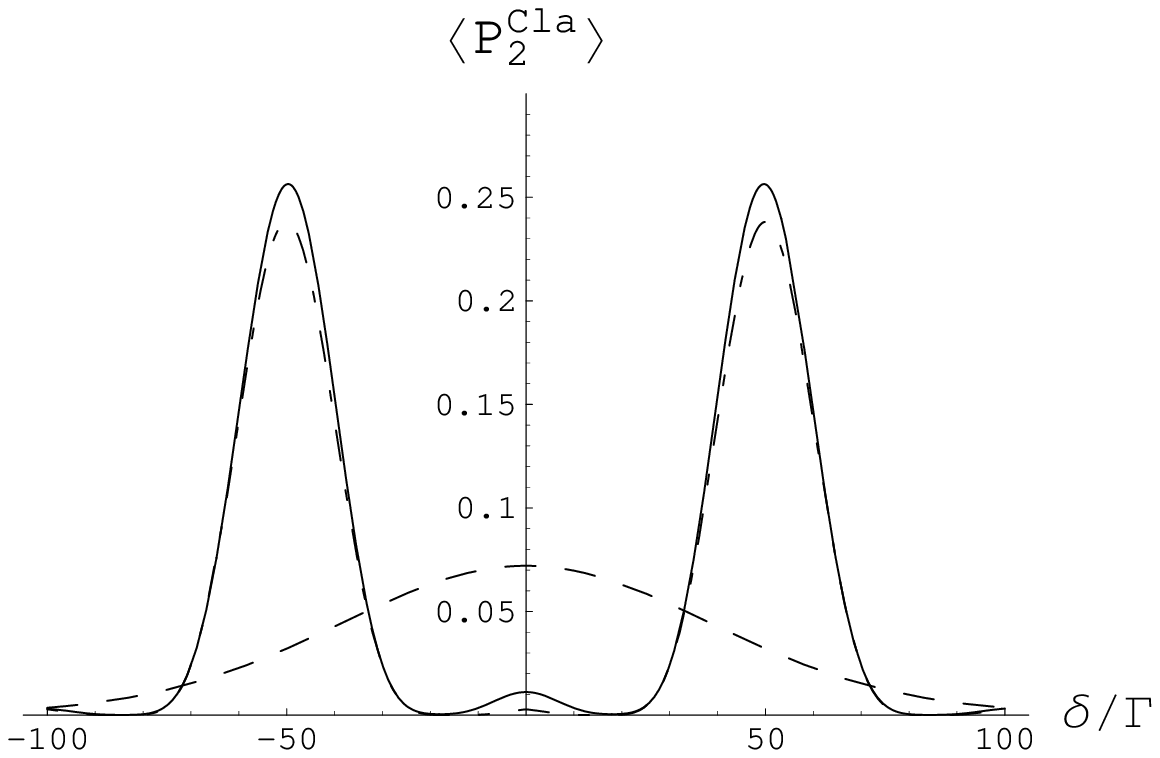}\textbf{c.}\\
\includegraphics[height=1.4in, width=2in]{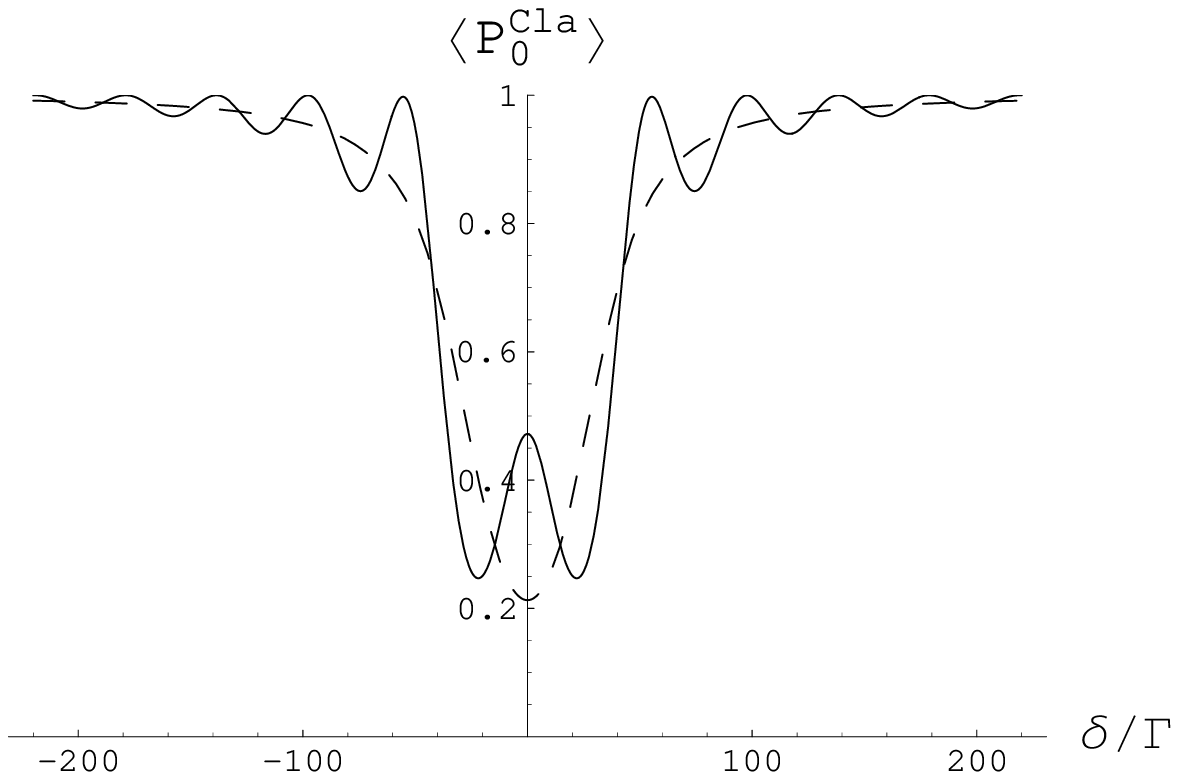}\textbf{d.}&
\includegraphics[height=1.4in, width=2in]{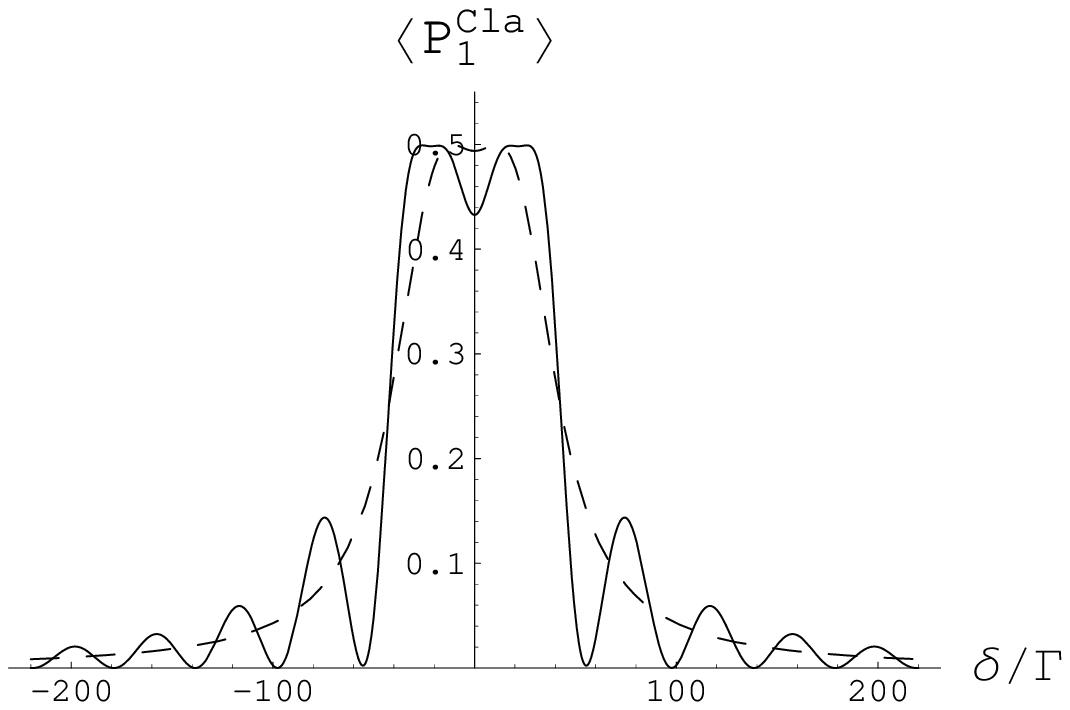}\textbf{e.}&
\includegraphics[height=1.4in, width=2in]{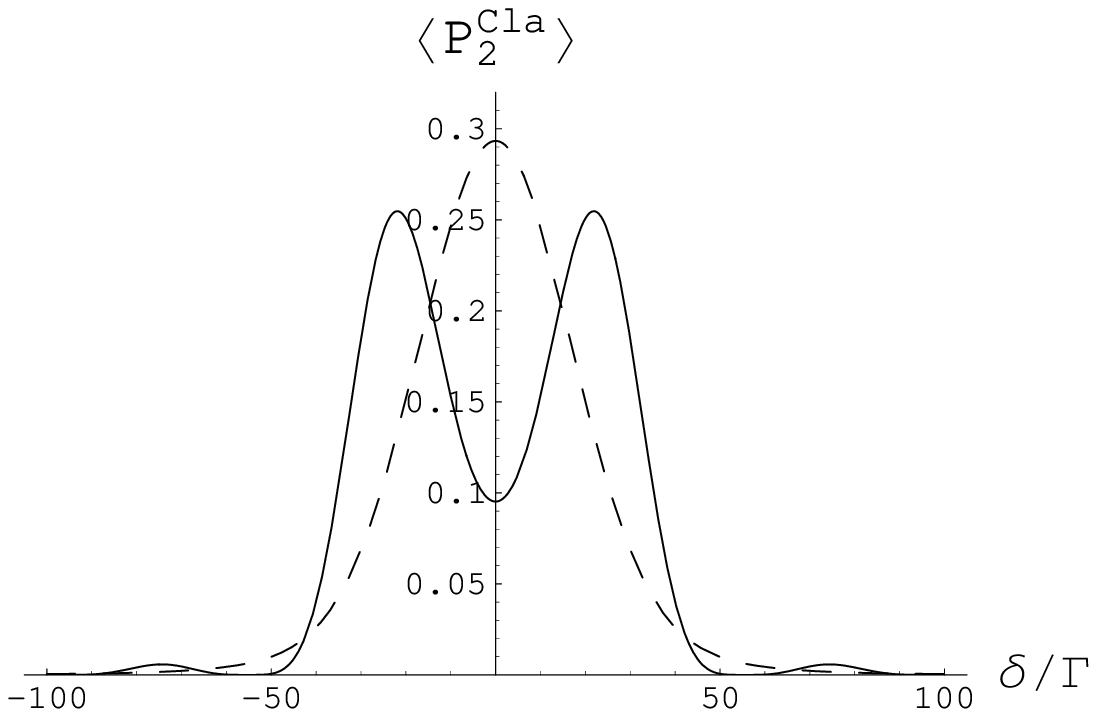}\textbf{f.}\\
 \includegraphics[height=1.4in, width=2in]{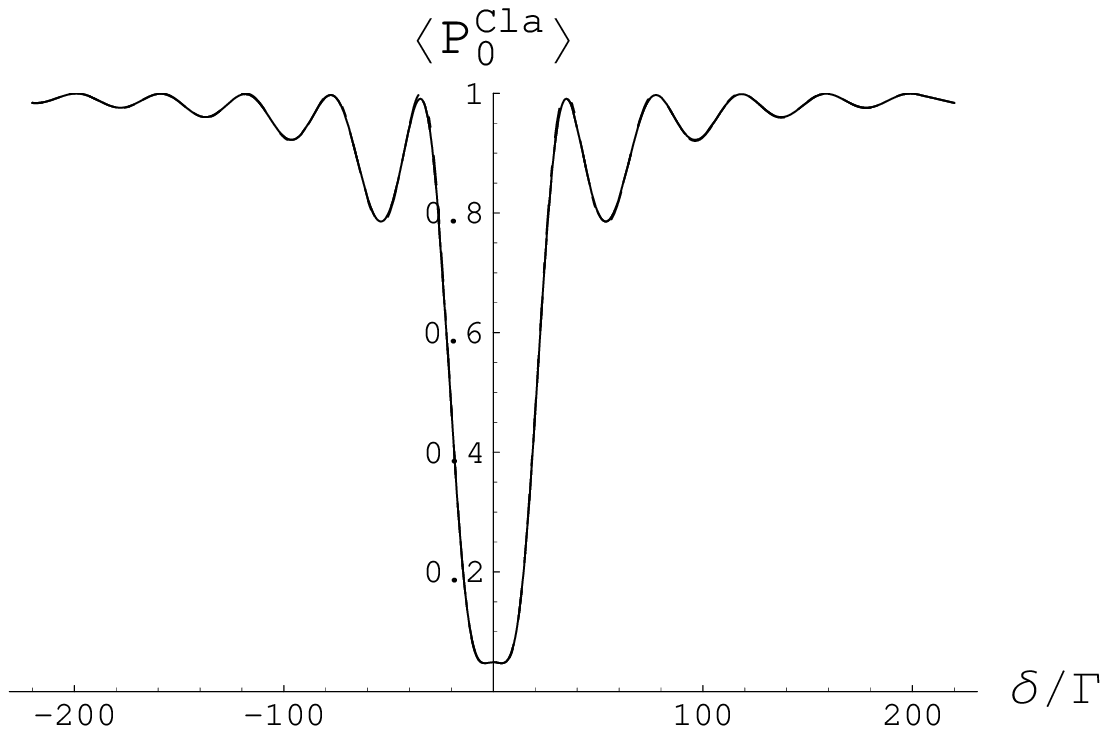}\textbf{g.}&
 \includegraphics[height=1.4in,width=2in]{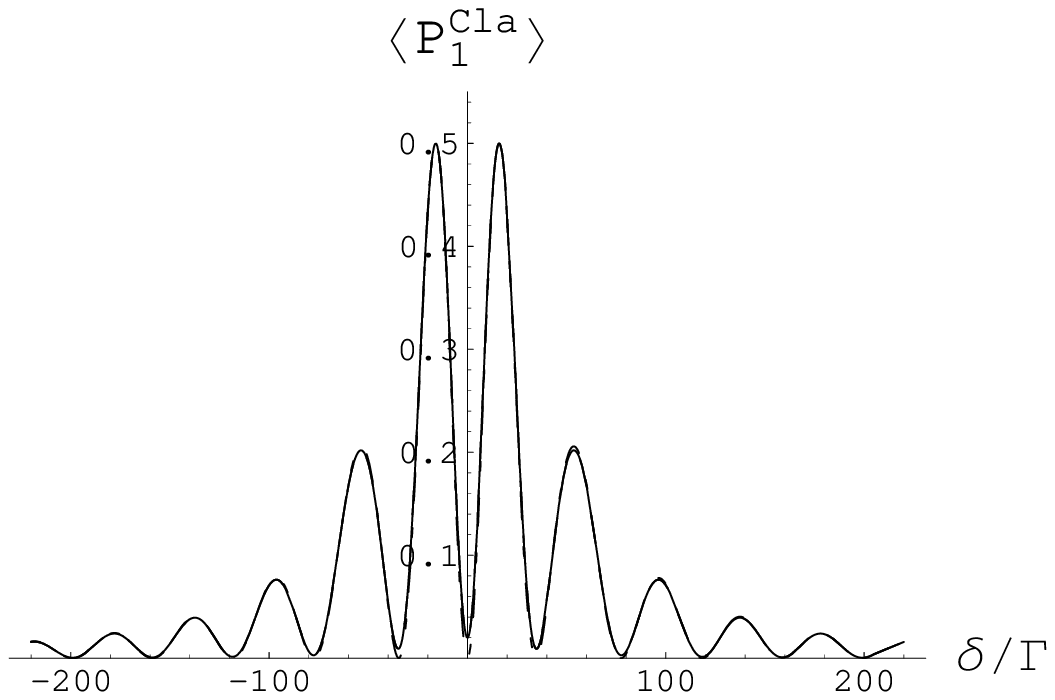}\textbf{h.}&
\includegraphics[height=1.4in, width=2in]{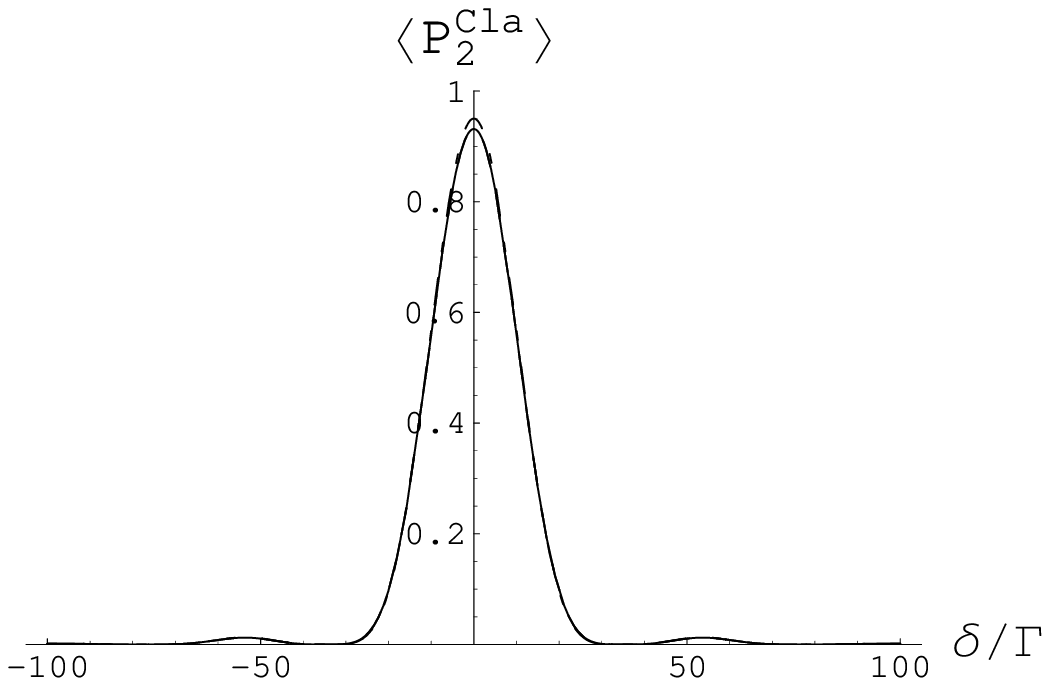}\textbf{i.}
\end{eqnarray*}
\caption{Photon statistics for two identical $\pi$-pulses, where
$\Delta=3\Gamma^{-1}$, $R=\Gamma=1$. {\em In the first row}: the
probabilities of emitting \textbf{a.}~0~\textbf{b.}~1 ~and
~\textbf{c.}~2~ photons in the case of the selective limit
$\nu=50\Gamma, ~~\Omega=20\Gamma$. The exact solution for the
two-state Kubo-Anderson process Eqs.~(\ref{p1av}-\ref{p0av}) is
represented by the smooth curves. The dot-dashed line represents the
semiclassical selective limit approximation
Eqs.~(\ref{p1avsel}-\ref{p0avsel}) for the two state process. And
the dashed curves show the Gaussian process.; ~~{\em In the second
row}: the probabilities of emitting \textbf{d.}~0~\textbf{e.}~1 ~and
~\textbf{f.}~2~ photons in the intermediate situation $\nu=20\Gamma,
~~\Omega=20\Gamma$.;~~ {\em In the third row}: the probabilities of
emitting \textbf{g.}~0~\textbf{h.}~1 ~and ~\textbf{i.}~2~ photons in
the case of the impulsive limit $\nu=2\Gamma, ~~\Omega=20\Gamma$.
The dot-dashed line represents the semiclassical impulsive limit
approximation Eqs.~(\ref{PnImp1}-\ref{PnImp0}). (The curves
representing the exact solution for the telegraph and Gaussian
noises, and the impulsive limit approximation coincide). Pay
attention, the graphs are scaled differently.}
\end{figure}
\end{widetext}
When $\delta_L=\pm\nu$ from Eqs.~(\ref{p1avsel}-\ref{p0avsel}) we
find:
\begin{equation}\!\!\!\!\!\!\!\!\!\!\!\lim_{|\delta_L|=\nu\gg\Omega}\langle P_{1}^{\rm Cla} \rangle =
\frac{1+e^{-2R\Delta}}{8}\sin^2\left(\Omega T\right)+
\frac{1-e^{-2R\Delta}}{2}\sin^2\left(\frac{\Omega T}{2}\right),
\label{eqClSel1}\end{equation}
\begin{equation}\lim_{|\delta_L|=\nu\gg\Omega}\langle P_{2}^{\rm Cla}
\rangle
=\frac{(1+e^{-2R\Delta})(1-e^{-\Gamma\Delta})}{4}\sin^4\left(\frac{\Omega
T}{2}\right). \label{eqClSel2}\end{equation}
%
%
%
Note, that Eqs.~(\ref{eqClSel1},\ref{eqClSel2}) exhibit Rabi
oscillations. Applying Eqs.~(\ref{eqClSel1},\ref{eqClSel2}) to the
$\pi$-pulses gives:
\begin{equation}\lim_{|\delta_L|=\nu\gg\Omega}\langle P_1^{\rm Cla}
\rangle = {1 \over 2} \left( 1 - e^{ - 2 R \Delta} \right),
\label{P1selpi}\end{equation}
\begin{equation}
\lim_{|\delta_L|=\nu\gg\Omega}\langle P_2^{\rm Cla}\rangle ={1 \over
4} \left( 1 + e^{ - 2 R \Delta} \right) \left( 1 - e^{ - \Gamma
\Delta} \right),\label{P2selpi}\end{equation}
and
\begin{equation}
\lim_{|\delta_L|=\nu\gg\Omega}\langle P_0^{\rm Cla} \rangle = {1
\over 4} \left( 1 + e^{ - 2 R \Delta} \right) \left( 1 + e^{ -
\Gamma \Delta} \right). \label{P0selpi}\end{equation}
The results of Eqs.~(\ref{P1selpi}-\ref{P0selpi}) make perfect
physical sense. For example a single photon may be emitted only if
the absorption frequency $\omega(t)$ is found once in the $+$ state
and once in the $-$ state, since only one of these states is in
resonance with the laser. The system, which is in the ground state
at the beginning of the experiment, gets excited exactly once, and
nothing interrupts the spontaneous emission process. Hence, $\langle
P_1 \rangle = {\cal P}[+,-] + {\cal P}[-,+]$. Note, that for the
strong $\pi$-pulses in the case of the impulsive limit we have
$\langle P_1 \rangle = 0$ (see Eq.~(\ref{Eqpooo1})).\\

\section{Demonstration of results}

In Fig.~3 we plotted the semiclassical parts of $\langle
P_0\rangle$, $\langle P_1\rangle$ and $\langle P_2\rangle$ for the
two state and Gaussian processes for two identical $\pi$-pulses. We
see the transition from the selective limit (first row) to the
impulsive limit (third row), where the graphs corresponding to the
two processes visually coincide. The graphs in Fig.~3 clearly
provide the information on the spectral shifts $\nu$. Finally, it is
worth noticing, that the behavior of photon statistics corresponding
to the two-state process is oscillatory in $\delta_L$, while in the
case of the Gaussian noise it is not. The origin of this effect
follows from the fact, that the matrix elements of the propagators
${\cal G}\left[\omega\left(t_i\right)\right]$
Eqs.~(\ref{calG},\ref{matrxelm}), and hence the probability of n
photon emission events, $P_n$ are sinusoidal functions of
$\delta_L-\delta w(t)$. Therefore in the case of discrete dichotomic
noise the phase of the sinusoidal functions we are summing up takes
only two values $\pm\nu$. However, in the case of the continuous
Gaussian distribution we integrate over a continuous range of
phases. Thus, as we approach the selective limit in the case of
Gaussian noise, the oscillations in $\delta_L$ are destroyed by averaging.\\

In Fig.~4 we plotted the three dimensional (above) and contour
(below) graphs of $\langle P_1^{\rm Cla}\rangle$ for the two state
Kubo-Anderson process as a function of the delay interval $\Delta$
and the spectral shift $\nu$. The graphs were obtained for the case
of two $\pi$-pulses for a constant values of $\Omega=60\Gamma$,
$R=3\Gamma$, $\Gamma=1$ (hence, $RT\simeq 0.157$) in pure resonance
with the $+$ state $\delta_L=\nu$. These graphs clearly show, that
the photon statistics of the pump-probe set up yields the dynamical
information on the rate $R$. However, for $\nu\ll\Omega$ this
information cannot be obtained, since $\langle P_1^{\rm Cla}\rangle$
stays visibly unchanged along the $\Delta$-axes, as expected. For
larger shifts the changes along the $\Delta$-axes become more and
more significant, and $\langle P_1^{\rm Cla}\rangle$ can be used to
measure $R$.

\begin{figure}
 \includegraphics[height=4in, width=2.4in]{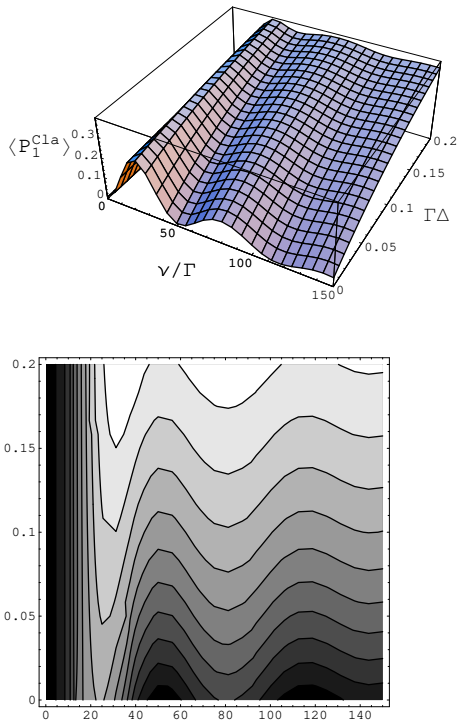}
\caption{The plot of $\langle P_1^{\rm Cla}\rangle$ for the two
state Kubo-Anderson process as a function of the delay $\Delta$ and
spectral shift $\nu$ in the case of pure resonance with the $+$
state $\delta_L=\nu$ for two identical $\pi$-pulses with
$\Omega=60\Gamma$, $R=3\Gamma$. The lower graph is the contourplot
of the upper three-dimensional graph.}
\end{figure}
%

\subsection{Measurement Limitations}

Combining the requirements $\Omega T\sim 1$ and $RT\ll 1$ implies
$\Omega\gg R$. Hence, only the three following limiting cases may
occur:
\begin{eqnarray*}
(a)~\Omega\gg\nu\gg R& -& \mbox{impulsive pulse, slow process}\\
(b)~\Omega\gg R\gg\nu& -& \mbox{impulsive pulse, fast process}\\
(c)~\nu\gg\Omega\gg R& -& \mbox{selective pulse, slow
process,}\end{eqnarray*}
while all the other combinations violate our
assumptions.\\

As mentioned, the direct dependence of the photon statistics on the
shift $\nu$ and the rate $R$ becomes undetectable when the strength
of the laser field reaches the impulsive limit. We saw (see Fig.~2),
that for $\nu\leq\Omega$ the information on the stochastic process
encoded in photon statistics becomes equivalent to the information
provided by line-shape in continuous wave spectroscopy. Since
$\Omega\gg R$ we see, that the effectiveness of the pump-probe
technique is restricted to slow processes $\nu> R$. Thus, the
assumptions $RT\ll 1$ and $\Omega T\sim 1$ determine the limitations
of the measurement of stochastic dynamics, and to get information on
wider range of dynamics our method must be improved by removing
these conditions. In our future work we plan to focus on the
investigation of processes,
where $RT\ll1$ is not fulfilled.\\

%
\section{Fast Modulation} Finally, let us consider an interesting
case of the fast modulation limit $R\to\infty$, $\nu\to\infty$ such
that $R\gg \nu$, where the motional narrowing effects take place.
From inequalities $(a,b,c)$ follows, that the fast modulation limit
must be impulsive limit as well. Therefore,
Eqs.~(\ref{Eqpooo0}-\ref{Eqpooo2}) hold with
\begin{equation}
\lim_{R\gg\nu}\phi(\Delta) =\exp( - \Gamma_{{\rm SD}} \Delta/2),
\label{CorGausKubo} \end{equation}
with
\begin{equation}
\Gamma_{SD}=\nu^2/R,\label{GammaSD}\end{equation}
which means that $(\Gamma_{{\rm
SD}} + \Gamma)/2$ is the renormalized decay rate which damps the
coherent terms. Note, that $\Gamma_{SD}$ is a measurable physical
observable determining the width of the line shape in continuous
wave spectroscopy \cite{Kubo}. \\

  The results of Eq.~(\ref{CorGausKubo},\ref{GammaSD}) are obtained
straightforwardly for the two state Kubo-Anderson process using our
exact solution Eqs.~(\ref{eqPn13F1},\ref{eqPn13F2}). We calculate
the limit $R\gg\nu$ in such a way, that $\Gamma_{{\rm SD}}=\nu^2/R$
is finite. For example considering Eq.~(\ref{eqPn13F2}) we have:
%
%
$$
\hat{h}\left[\mp,\pm,-2 i \nu\right] = e^{ - \Delta( R - i \nu) } {
\sinh\left[ \Delta \sqrt{ R^2 - \nu^2} \right] R \over 2 \sqrt{ R^2
- \nu^2}}\simeq$$\begin{equation} \simeq\frac{e^{ - \Delta( R - i
\nu) }}{4} \left( e^{ R\Delta \sqrt{ 1 - \frac{\nu^2}{R^2}}} -e^{-
R\Delta \sqrt{ 1 - \frac{\nu^2}{R^2}}}\right)\label{temp0}
\end{equation}
%
Using $\sqrt{1-\frac{\nu^2}{R^2}}\sim1-\frac{\nu^2}{2R^2}$ in the
exponent of Eq.~(\ref{temp0}) we obtain:
%
%
%
$$
\lim_{R\gg\nu}\hat{h}(\mp,\pm,-2 i \nu)=\lim_{R\gg\nu}\frac{e^{ -
\Delta( R - i \nu) }}{4} \left( e^{ R\Delta (1-\frac{\nu^2}{2R^2})}
-\right.$$
\begin{equation}\left.-e^{- R\Delta (1-\frac{\nu^2}{2R^2})}\right)=\frac{1}{4} e^{
i\Delta\nu }\exp(-\frac{\Delta}{2}\frac{\nu^2}{R}),\label{fastmod}
\end{equation}
provided $R\Delta\gg1$. Exactly the same procedure applied to
Eq.~(\ref{eqPn13F1}) leads to similar result. Inserting the result
Eq.~(\ref{fastmod}) into Eq.~(\ref{eqPn11}) cancels the oscillating
term $e^{ i\Delta\nu }$. Therefore, the stochastic phase in the
limit of fast modulation reduces to $\exp\left(-\frac{\Delta}{2}
\frac{\nu^2}{R}\right)$. It is easy to show, that
Eq.~(\ref{CorGausKubo}) is valid as well for the Gaussian process
under investigation, with $\Gamma_{SD}=\nu^2/2R$. The result
Eqs.~(\ref{CorGausKubo}) obtained in this section for the telegraph
noise exhibits the fast modulation limit behavior general
for all Markovian processes \cite{Kubo}.\\

%
\section{Summary}

 Theoretical investigation of the new field of
single molecule non-linear spectroscopy was presented. We have
obtained analytically the exact expressions for photon statistics
emerging from the interaction of the pump-probe set up with a single
two level system in terms of quantum trajectories. The theory
clearly emphasized two types of terms: the coherent and the
semiclassical, the later independent on the phase accumulated
by the coherences during the delay interval between the pulses.\\

The following conclusions were made:\\

1. Unlike the line-shape in continuous wave experiments, the photon
statistics obtained from pump-probe set up depends not only on the
spectral shifts $\nu$, but also on the rate $R$ of the spectral
diffusion process and exhibits oscillations not found in a
line-shape.\\

2. In the limit of the impulsive pulses the spectral selectivity is
lost. The information on the spectral diffusion parameters,
contained only in the phase accumulated during the delay interval,
is then equivalent to the information gained from a line-shape in
continuous wave spectroscopy. In optics using pump-probe technique
this information is expected to be difficult to detect, because of
the huge bare optical frequency $\omega_0$ and dephasing effects.
It will be interesting to investigate the coherence effect in multilevel systems.\\

3. In contrast, in the selective limit, where the laser field is
weak compared to the deviations of the spectral shifts, the photon
statistics offers full information on spectral diffusion parameters.\\

4. We have shown, that the limitation of the measurement of
stochastic dynamics is determined by the conditions $RT\ll 1$ and
$\Omega T\sim 1$, restricting the effectiveness of pump-probe set up
only to slow processes $\nu\gg R$. Hence, to get better results our
method must be improved by removing these conditions. In particular,
we plan to investigate photon statistics for processes with
$RT\sim1$. In addition, our techniques can be easily modified for
the investigation of relaxation processes \cite{vanDijk}, where the
frequency shift may be large, thus
allowing the measurement of dynamics up to the order of pico-seconds.\\

5. Finally, in the fast modulation limit where the stochastic
transitions rate $R$ is much larger then spectral shifts $\nu$, the
influence of the underlying spectral diffusion dynamics on the
photon statistics reduces to the well-known Kubo-Anderson
correlation function coefficient $\exp\left(-\frac{\Delta}{2}
\Gamma_{SD}\right)$, which damps the
contribution of the coherent terms.\\

$$\mbox {\textsc{acknowledgment}}$$This work was supported by the
Israel Science Foundation.\\

%
\section{Appendix A}\label{appendixA} 

{\em Photon Statistics.}~~~~Starting with \cite{Zoller,Mollow} an
interpretation of the optical Bloch formalism yields a tool for the
calculation of photon statistics. The formal solution to
Eq.~(\ref{eqBloch}) may be given by the infinite iterative expansion
in $\hat{\Gamma}$ \cite{Mukamel}:
$$ \sigma_{(t)} = {\cal G}(t,0) \sigma_{(0)} + \int_0 ^t {\rm d t_1} {\cal G} (t,t_1) \hat{\Gamma} {\cal G}(t_1, 0) \sigma_{(0)} + $$
\begin{equation}
+\int_0 ^{t} {\rm d} t_2 \int_0 ^{t_2} {\rm d} t_1 {\cal G}( t, t_2)
\hat{\Gamma} {\cal G}(t_2 , t_1) \hat{\Gamma} {\cal G} (t_1 , 0 )
\sigma_{(0)} + \cdots, \label{eqformal}
\end{equation}\\
where $\sigma_{(0)}$ is the initial condition, and the Green
function describing the evolution of the system in the absence of
spontaneous transitions into the ground state (i.e. without
$\hat{\Gamma}$ ) is
\begin{equation}
{\cal G} (t,t') = \hat{T} \exp\left[ \int_{t'} ^ t L(t_1) {\rm d}
t_1 \right], \label{eqGreenf}
\end{equation}
where $\hat{T}$ is the time ordering operator.\\

Each term in the expansion Eq.~(\ref{eqformal}) describes the
propagation of the system emitting exact number of photons: for
example the first term does not include $\hat{\Gamma}$ at all, and
describes a process where no photons are emitted, the second term
corresponds to the processes where only one photon is emitted and so
on. Therefore, $\sigma^{(n)}_{(t)}$ defined as:
\begin{equation}
 \sigma^{(n)}_{(t)} = U^{(n)}_{(t,0)} \sigma_{(0)}
\label{eqsign}
\end{equation}
describes the conditional state of the system at the moment $t$,
provided that n photon emission events occurred in the time interval
$(0,t)$, and
\begin{equation}
 U^{(n)}_{(t,t')} =
\int_{t'} ^t {\rm d} t_n \cdots \int_{t'} ^{t_2} {\rm d} t_1~ {\cal
G}(t,t_n) \hat{\Gamma} \cdots\hat{\Gamma} {\cal G}(t_1,t')
\label{eqUn}
\end{equation}
is called the n-photon-propagator.\\

The main equation for calculating the probability of $n$ emission
events up to time $t$ is:
\begin{equation}
P_n(t) = (\langle {\rm e}|+\langle {\rm g}|)\sigma^{(n)}_{(t)}
\rangle=(\langle {\rm e}|+\langle {\rm
g}|)U^{(n)}_{(t,0)}|\sigma_{(0)}\rangle, \label{eqPn}
\end{equation}
which is simply the trace of the density matrix conditioned by $n$
emission events. We showed in \cite{SHB} that in the case of two
separated pulses the total n-photon-propagator acting from $t=0$ up
to $t$ may be written as:
\begin{equation}
\begin{array}{c}
 U^{(n)}_{(t,0)}=U^{(n-\alpha-\beta-\gamma)}_{(t,t_3)}U^{(\gamma)}_{(t_3,t_2)}U^{(\beta)}_{(t_2,t_1)}U^{(\alpha)}_{(t_1,0)},\end{array}
\label{eqSumUU}
\end{equation}
where the superscripts $\alpha$, $\beta$ and $\gamma$ are all
non-negative integers leading to $n$ emission events (i.e.~
$n-\alpha-\beta-\gamma\geq0$). The Einstein's summation rule from 0
to n must be applied to every superscript appearing twice.
Eq.~(\ref{eqSumUU}) means that the n-photon-propagator acting in
$(0,t)$ can be decomposed into the sum of all possible products of
the $\alpha$-photon-propagator acting in $(0,t_1)$,
$\beta$-photon-propagator acting in $(t_1,t_2)$,
$\gamma$-photon-propagator acting in $(t_2,t_3)$ and
$(n-\alpha-\beta-\gamma)$-photon-propagator acting in $(t_3,t)$. \\

The propagators acting during the delay interval and after the
second pulse, where the laser is off and $\Omega=0$, may be found
immediately:
%
 \begin{equation}U^{(0)}_{(t_1+\Delta,t_1)}
 =\left(
\begin{array}{c c c c}
e^{-\Gamma\Delta}  & 0 & 0  & 0 \\
 0 & 1 & 0  &  0 \\
 0 & 0  & e^{ i \omega_0\Delta - \frac{\Gamma\Delta}{2}}  & 0 \\
 0 & 0  & 0  & e^{ -(i \omega_0\Delta + \frac{\Gamma\Delta}{2})}
\end{array}\right),
\label{eqU0}
\end{equation}
%
%
\begin{equation} U^{(1)}_{(t_1+\Delta,t_1)}= \left(
\begin{array}{c c c c}
 0 & 0 & 0  & 0 \\
 1-e^{-\Gamma\Delta} & 0 & 0  &  0 \\
 0 & 0  & 0  & 0 \\
 0 & 0  & 0  & 0
\end{array}
\right), \label{eqU1}\end{equation}
%
and
\begin{equation}
U^{(n)}_{(t_1+\Delta,t_1)}=0\mbox{~~ for
$n>1$}\label{U}.\end{equation}
The propagators acting in $(t_3,t)$, where $t\rightarrow\infty$, are
given by the limit $\Delta\rightarrow\infty$.\\

To obtain photon statistics for short pulses in the summation
Eq.~(\ref{eqSumUU}) we pick up only those processes, where the
propagators acting during the pulse events are the
zero-photon-propagators, i.e. $U^{(0)}_{(t_1,0)}={\cal
G}[\omega(t_1)]$ and $U^{(0)}_{(t_3,t_2)}={\cal G}[\omega(t_2)]$. We
substitute Eqs.~(\ref{eqSumUU}, \ref{eqU0}, \ref{eqU1}) into
Eq.~(\ref{eqPn}), and  inserting the closure relation $
\sum_{j=e,g,c,c^*}|j \rangle\langle j|=1 $ between each two
propagators obtain
Eq.~(\ref{PnSD}) of the article.\\

The calculation of the matrix elements ${\cal G}[\omega(t_i)]$ is
made using RWA. Applying RWA to the optical Bloch
equation~(\ref{eqBloch}) consists of neglecting fast oscillating
non-resonant terms \cite{CT}. As a result the following equation is
obtained:
\begin{equation}
\dot{\sigma}^{\rm RWA} = L^{\rm RWA}(t) \sigma^{\rm RWA} +
\hat{\Gamma}\sigma^{\rm RWA},\label{eqBloch1}
\end{equation}
where
%
%
\begin{equation}
L^{\rm RWA}(t) = \left(
\begin{array}{c c c c}
-\Gamma  & 0 & {-i \Omega \over 2} & {i \Omega \over 2} \\
0 & 0 & {i \Omega \over 2} & {-i \Omega \over 2} \\
 {-i \Omega \over 2} & {i \Omega \over 2} & - {\Gamma\over 2} - i \delta(t) & 0 \\
 {i \Omega \over 2} & {-i \Omega \over 2} & 0 & - {\Gamma\over 2} + i
 \delta(t)
\end{array}
\right), \label{eqLRWA}
\end{equation}
and $\delta(t)$ Eq.~(\ref{detuning}) is the detuning at moment t. As
discussed in the article, the detuning is assumed to be constant
during the pulse.
%
%
Hence, in the new representation the calculation of the Green
function is simple:
\begin{equation}
 {\cal G}[\omega(t_i)]= \exp\left[T L^{\rm RWA}[\omega(t_i)]
\right].\label{U0}
\end{equation}
Since the interaction time goes to zero $\Gamma T\to 0$, the
spontaneous emission effects during the pulses are completely
suppressed, and we find:
%
\begin{widetext}
%
%
$$
\langle {\rm e}|{\cal G}[\omega(t_i)]|{\rm g}\rangle=\langle {\rm
g}|{\cal G}[\omega(t_i)]|{\rm e}\rangle=1-\langle {\rm g}|{\cal
G}[\omega(t_i)]|{\rm g}\rangle =1-\langle {\rm e}|{\cal
G}[\omega(t_i)]|{\rm e}\rangle
=\frac{\Omega^2\sin^2\left[\frac{\Omega
T}{2}\sqrt{1+\frac{(\delta_L-\delta
w(t_j))^2}{\Omega^2}}\right]}{(\delta_L-\delta w(t_j))^2+\Omega^2}$$
$$
\langle {\rm c}|{\cal G}[\omega(t_i)]|g\rangle=\langle {\rm
c}^{*}|{\cal G}[\omega(t_i)]|g\rangle^{*}=\langle {\rm g}|{\cal
G}[\omega(t_i)]|{\rm c}\rangle=\langle {\rm g}|{\cal
G}[\omega(t_i)]|{\rm c}^{*}\rangle^{*}=$$ $$=-\langle {\rm c}|{\cal
G}[\omega(t_i)]|{\rm e}\rangle=-\langle {\rm c}^{*}|{\cal
G}[\omega(t_i)]|{\rm e}\rangle^{*}=-\langle {\rm e}|{\cal
G}[\omega(t_i)]|{\rm c}\rangle= -\langle {\rm e}|{\cal
G}[\omega(t_i)]|{\rm c}^{*}\rangle^{*} =$$
\begin{equation}=\frac{\Omega\left(2(\delta_L-\delta w(t_j))
\sin^2\left[\frac{\Omega T}{2}\sqrt{1+\frac{(\delta_L-\delta
w(t_j))^2}{\Omega^2}}\right]+i\sqrt{(\delta_L-\delta
w(t_j))^2+\Omega^2}\sin\left[\Omega T\sqrt{1+\frac{(\delta_L-\delta
w(t_j))^2}{\Omega^2}}\right]\right)}{2\left[(\delta_L-\delta
w(t_j))^2+\Omega^2\right]}. \label{matrxelm}
\end{equation}
%
%
For pulses satisfying $\frac{\delta_L-\delta w(t_i)}{\Omega}\ll1$,
i.e. resonant or impuslive pulses, from Eq.~(\ref{matrxelm}) we
find:
\begin{equation}
{\cal G}[\omega(t_i)] = \left(
\begin{array}{c c c c}
  \cos^2 {\Omega T \over 2}  &  \sin^2 { \Omega T \over 2}   & - i {\sin \Omega T  \over 2}  & i {\sin \Omega T \over 2} \\
 \sin^2 { \Omega T \over 2}  & \cos^2 { \Omega T \over 2}   &  i {\sin \Omega T  \over 2}  & - i  {\sin \Omega T \over 2} \\
 -i {\sin \Omega T \over 2} & i {\sin \Omega T \over 2}  &  \cos^2 {\Omega T \over 2}  &  \sin^2 { \Omega T \over 2}  \\
 i {\sin \Omega T \over 2} & - i  {\sin \Omega T \over 2}  &  \sin^2 {\Omega T \over 2} &  \cos^2 { \Omega T \over 2}
\end{array}
\right). \label{RabiOscl}
\end{equation}
Substituting $\Omega T=\pi$ into Eq.~(\ref{RabiOscl}) we see that
the off-diagonal terms giving rise to coherent terms vanish. More
detailed discussion on RWA and its application to the
calculation of the matrix element may be found in \cite{SHB}.\\
%
%

\section{Appendix B} \label{Gaussian}

{\em Semiclassical approximation for the two state Kubo-Anderson
process.}~~~~Here we present the exact expressions for $\langle
P_n^{\rm Cla}\rangle$ (n=0,1,2) for the case of the two state
Kubo-Anderson process. Using Eqs.~(\ref{Pif}), Table~1, the matrix
elements Eqs.~(\ref{matrxelm}) and Eq.~(\ref{PnSDgen}) we find:
$$\langle P_1^{{\rm
Cla}}\rangle=\frac{(1+e^{-2R\Delta})}{2}\left\{\frac{\Omega^2\sin^2\left[\frac{\Omega
T}{2}\sqrt{1+\frac{(\delta_L-\nu)^2}{\Omega^2}}\right]}{(\delta_L-\nu)^2+\Omega^2}\times\frac{(\delta_L-\nu)^2+\Omega^2\cos^2\left[\frac{\Omega
T}{2}\sqrt{1+\frac{(\delta_L-\nu)^2}{\Omega^2}}\right]}{(\delta_L-\nu)^2+\Omega^2}+\right.$$
$$\left.+\frac{\Omega^2\sin^2\left[\frac{\Omega
T}{2}\sqrt{1+\frac{(\delta_L+\nu)^2}{\Omega^2}}\right]}{(\delta_L+\nu)^2+\Omega^2}\times\frac{(\delta_L+\nu)^2+\Omega^2\cos^2\left[\frac{\Omega
T}{2}\sqrt{1+\frac{(\delta_L+\nu)^2}{\Omega^2}}\right]}{(\delta_L+\nu)^2+\Omega^2}\right\}
$$
$$+\frac{(1-e^{-2R\Delta})}{2}\left\{\frac{\Omega^2\sin^2\left[\frac{\Omega
T}{2}\sqrt{1+\frac{(\delta_L-\nu)^2}{\Omega^2}}\right]}{(\delta_L-\nu)^2+\Omega^2}\times\frac{(\delta_L+\nu)^2+\Omega^2\cos^2\left[\frac{\Omega
T}{2}\sqrt{1+\frac{(\delta_L+\nu)^2}{\Omega^2}}\right]}{(\delta_L+\nu)^2+\Omega^2}+\right.$$
\begin{equation}
\left.+\frac{\Omega^2\sin^2\left[\frac{\Omega
T}{2}\sqrt{1+\frac{(\delta_L+\nu)^2}{\Omega^2}}\right]}{(\delta_L+\nu)^2+\Omega^2}\times\frac{(\delta_L-\nu)^2+\Omega^2\cos^2\left[\frac{\Omega
T}{2}\sqrt{1+\frac{(\delta_L-\nu)^2}{\Omega^2}}\right]}{(\delta_L-\nu)^2+\Omega^2}\right\},\label{p1av}
\end{equation}
$$
\langle P_2^{{\rm Cla}}\rangle
=\frac{(1+e^{-2R\Delta})(1-e^{-\Gamma\Delta})}{4}\left\{\frac{\Omega^4\sin^4\left[\frac{\Omega
T}{2}\sqrt{1+\frac{(\delta_L-\nu)^2}{\Omega^2}}\right]}{\left((\delta_L-\nu)^2+\Omega^2\right)^2}+
\frac{\Omega^4\sin^4\left[\frac{\Omega
T}{2}\sqrt{1+\frac{(\delta_L+\nu)^2}{\Omega^2}}\right]}{\left((\delta_L+\nu)^2+\Omega^2\right)^2}\right\}+$$
\begin{equation}
+\frac{(1-e^{-2R\Delta})(1-e^{-\Gamma\Delta})}{2}\times\frac{\Omega^4\sin^2\left[\frac{\Omega
T}{2}\sqrt{1+\frac{(\delta_L-\nu)^2}{\Omega^2}}\right]}{(\delta_L-\nu)^2+\Omega^2}
\times\frac{\sin^2\left[\frac{\Omega
T}{2}\sqrt{1+\frac{(\delta_L+\nu)^2}{\Omega^2}}\right]}{(\delta_L+\nu)^2+\Omega^2}.
\label{p2av}
\end{equation}
and
\begin{equation}
\langle P_0^{{\rm Cla}}\rangle =1-\langle P_1^{{\rm Cla}}\rangle
-\langle P_2^{{\rm Cla}}\rangle .\label{p0av} \end{equation}
The only approximation made in the semiclassical
Eqs.~(\ref{p1av}-\ref{p0av}) is neglecting the spontaneous emission
effects during the pulse events and applying RWA within the
calculation of the matrix elements. Since these expressions involve
the both $\nu$ and $R$, they may be used for the determination of
the spectral shifts and rates.
\end{widetext}
%

%
\section{Appendix C} \label{2sKA}



{\em Derivation of
$\hat{h}\left[\omega(t_1),\omega(t_2),-2i\nu\right]$.}~~~~The
two-state Kubo-Anderson process has been investigated extensively by
many authors, in particular different techniques for calculation of
the occupation times were proposed \cite{Berez,Luck}. Here we follow
the method used in \cite{Luck}. Providing the molecule was in the
$+$ state in the beginning of the first pulse and in the $-$ state
in the beginning of the second, so that exactly $m=2k+1$ where $k=0,
1, 2\cdots$ jumps occurred, it is possible to show \cite{Luck}, that
the double Laplace transform of $f_{m,\Delta}(T^+)$, the PDF of the
occupation time in the upper state $T^+$ for a fixed $\Delta$, is
given by:
\begin{equation}
\hat f_{m^{odd},s}(u)
=\hat\chi^{k+1}(s+u)\hat\chi^{k}(s)\frac{1-\hat\chi(s)}{2s},
\label{temp00}
\end{equation}
where
\begin{equation}
\hat\chi(s+u)=\int_0^{\infty}e^{-\tau(s+u)}\chi(\tau)\,d\tau, \;\;\;
\hat\chi(u)=\int_0^{\infty}e^{-\tau
u}\chi(\tau)\,d\tau\label{psiTrans}
\end{equation}
are Laplace transforms of ${\chi(\tau})$ - the PDF of the waiting
times between the subsequent jumps. The factor 2 in the nominator of
Eq.~(\ref{temp00}) is due to the probability to find the molecule in
the $+$ state in the beginning of the first pulse, which is equal
$1\over 2$ because there is symmetry between the states and we
assumed stationary process. Summing up $k$ from $0$ to $\infty$ we
have:
\begin{equation}
\hat{h}_s\left[+,-,u\right]
=\frac{1-\hat\chi(s)}{s}\frac{\hat\chi(s+u)}{1-\hat\chi(s+u)\hat\chi(s)},
\label{temp}
\end{equation}
where $\hat{h}_s(+,-,u)$ is the double Laplace transform of the PDF
of the occupation time in the $+$ state, providing the system was in
$+$ state in the beginning of the first pulse and in the $-$ state
in the beginning of the second independently of the number of
jumps.\\

Now applying Eq.~(\ref{temp}) to the Poissonian process:
\begin{equation}
\chi(\tau)=R e^{- R \tau}, \label{Poisson}
\end{equation}
which corresponds to the correlation function $\psi(\tau)$ given by
Eq.~(\ref{psi}), and substituting $u=-2i\nu$ we find:
\begin{equation}
\hat{h}_s\left[+,-,-2i\nu\right] =\frac{R}{2s(2R+s)-4i(R+s)\nu}.
\label{temp01}
\end{equation}
Finally the inverse Laplace transform $s\rightarrow\Delta$ leads to
Eq.~(\ref{eqPn13F2}) of the article:
$$
\hat{h}\left[+,-,-2 i \nu\right] = e^{ - \Delta( R - i \nu) } {
\sinh\left[ \Delta \sqrt{ R^2 - \nu^2} \right] R \over 2 \sqrt{ R^2
- \nu^2}}.
$$
Due to the symmetry of the stochastic process $\hat{h}\left[+,-,-2 i
\nu\right]=\hat{h}\left[-,+,-2 i \nu\right]$. The procedure of
derivation of Eq.~(\ref{eqPn13F1}) for $\hat{h}\left[\pm,\pm,-2 i
\nu\right]$ is similar.\\


Note, the marginal probabilities ${\cal
P}\left[\omega(t_1),\omega(t_2)\right]$ of finding the molecule in
state $\omega(t_1)=\pm$ during the first pulse and in state
$\omega(t_2)=\pm$ during the second pulse, independently of the
values of occupation times, are easy to derive from
Eqs.~(\ref{eqPn13F1},\ref{eqPn13F2}) by setting $\nu=0$ which is
equal to integrating out the $T^+$ from
$h\left[\omega(t_1),\omega(t_2),T^{+}\right]$. Thus, the following
obvious expressions are obtained:
$$ {\cal P}\left[\pm \pm\right]=\hat{h}\left[\pm,\pm,0\right]=\frac{1+\exp(- 2 R \Delta)}{4}
$$
and
$$
{\cal P}\left[\pm
\mp\right]=\hat{h}\left[\mp,\pm,0\right]=\frac{1-\exp(- 2 R
\Delta)}{4}.
$$
%

%


\begin{thebibliography}{99}
\bibitem{Remark} Originally, the contribution of the semiclassical paths to the
probability of one photon emission is: $P_1^{\rm Cla}=\langle {\rm
g} | {\cal G}\left[\omega(t_2)\right] | {\rm g} \rangle\langle {\rm
e} | {\cal G}\left[\omega(t_1)\right]| {\rm g} \rangle \left( 1 -
e^{ - \Gamma \Delta} \right) + \langle {\rm e} | {\cal
G}\left[\omega(t_2)\right]| {\rm e} \rangle \langle {\rm e} | {\cal
G} \left[\omega(t_1)\right] | {\rm g} \rangle e^{ - \Gamma \Delta}
$. However, due to the symetry of the matrix elements
Eqs.~(\ref{matrxelm}) the trajectories multiplied by
$e^{-\Gamma\Delta}$ cancel each other.
\end{thebibliography}
\end{document}